\documentstyle[epsfig, 12pt]{article}
\begin{document}

\begin{titlepage}

\hfill{October 1997}

\hfill{TMUP-HEL-9710}

\hfill{UM-P-97/57}

\hfill{RCHEP-97/14}

\vskip 2.0 cm

\centerline{{\large \bf 
Up-down atmospheric neutrino flux asymmetry predictions}} 
\centerline{{\large \bf for various neutrino oscillation scenarios}}
\vskip 1.0 cm
\centerline{R. Foot$^a$, R. R. Volkas$^a$ and O. Yasuda$^b$}
\vskip 0.7 cm
\noindent
\centerline{{\it $^a$ School of Physics,}}
\centerline{{\it Research Centre for High Energy Physics,}}
\centerline{{\it The University of Melbourne,}}
\centerline{{\it Parkville, 3052 Australia. }}
\vskip 0.5cm
\noindent
\centerline{{\it $^b$ Department of Physics, 
Tokyo Metropolitan University,}}
\centerline{{\it 1-1 Minami-Osawa Hachioji, Tokyo
192-03, Japan}}

\vskip 1.0cm

\centerline{Abstract} 

\vspace{0.7cm} 

\noindent 
We compute up-down asymmetries for atmospheric neutrino induced 
electron and muon events in the SuperKamiokande detector for the 
following neutrino oscillation models: (A)  maximally mixed 
$\nu_{\mu} - \nu_{\tau}$ or $\nu_{\mu}-\nu_s$, (B) maximally mixed 
$\nu_{\mu} - \nu_e$, (C) threefold maximal mixing between 
$\nu_e$, $\nu_{\mu}$ and $\nu_{\tau}$, and (D) neutrino oscillations 
via Equivalence Principle or Lorentz invariance violation. 
We emphasise the role of different momentum cuts.

\end{titlepage} 

The solar \cite{solar} and atmospheric \cite{kamioka1}-\cite{soudan2}
neutrino observations and the LSND \cite{lsnd} experiment provide strong
evidence that neutrinos have nonzero masses and oscillate. Many specific
neutrino mass scenarios have been proposed as explanations for some or all of
these results. These explanations will be put to stringent tests in the next
few years as more data are collected. 

In this paper we will focus on ways to experimentally test several different
proposed explanations of the atmospheric neutrino anomaly.  One good way to
discriminate between these possible explanations is through the up-down
asymmetry, with respect to zenith angle, of the detected charged leptons in
experiments such as SuperKamiokande \cite{fvb}\cite{fvy}\cite{lfp}. 
This idea was
introduced in Ref.\cite{fvb} and studied in detail in Ref.\cite{fvy} in the
context of the Exact Parity Model solution to the atmospheric neutrino
anomaly \cite{epm}. The up-down asymmetry idea was independently considered
by the
authors of Ref.\cite{lfp} in the case of six possible neutrino oscillation
models.  As emphasised in Ref.\cite{lfp}, the sign and energy dependence of
the up-down asymmetry will provide a useful diagnostic tool in
differentiating the various explanations of the atmospheric neutrino anomaly.
Reference \cite{lfp} focussed on the energy dependence of the up-down
asymmetry. A complementary approach is to study the asymmetry with various
cuts in momentum \cite{fvy}\cite{fn}. This is useful because the
Kamiokande and SuperKamiokande collaborations have divided their event sample
into two classes (sub-GeV and multi-GeV) on the basis of momentum cuts. In
the future, improved statistics may lead to the event sample being further
subdivided with respect to momentum. The purpose of this paper is to study
the up-down asymmetry expected at SuperKamiokande with various cuts in
momentum for a variety of neutrino oscillation models. 

It is {\it a priori} interesting to consider various cuts in zenith angle as
well as energy
when analysing the up-down asymmetries of charged lepton events in
SuperKamiokande \cite{fvy}. A class of measures of the up-down asymmetry
can be defined by introducing 
the quantities $y_e^{\eta}$ and $y_{\mu}^{\eta}$, where
\begin{equation}
y^{\eta}_e \equiv {N_e^{-\eta} \over N_e^{+\eta}},\qquad
y^{\eta}_\mu \equiv {N_\mu^{-\eta} \over N_\mu^{+\eta}}.
\end{equation}
Here $N_e^{-\eta}$ ($N_e^{+\eta}$) is the number of charged electrons
produced in
the detector with zenith angle $\cos \Theta < -\eta$ ($\cos \Theta > \eta$).
$N_{\mu}^{\pm\eta}$ are the analogous quantities for charged muons. The
zenith
angle $\Theta$ is defined so that $\Theta = 0$ corresponds to downward
travelling and $\Theta = \pi$ to upward
travelling charged leptons. Different choices for $\eta$ correspond to
different cuts in zenith angle. (Note that Ref.\cite{lfp} considered
the $\eta = 0$ case in detail for several different models, while
Ref.\cite{fvy} studied different choices for $\eta$ for the Exact Parity
Model.) Since $y^{\eta}_{e, \mu}$
are ratios of fluxes, they will be approximately free
of the systematic errors arising from uncertain absolute fluxes ($\pm
30\%$) as
well as uncertain $\mu/e$ flux ratios ($\pm$ a few percent).  
For symmetry reasons, $y^{\eta}_{e,\mu} \simeq 1$ 
is expected in the absence of neutrino oscillations.
If neutrino oscillations are present, however, then 
$y^{\eta}_{e, \mu} \neq 1$ is expected. 

In the calculations to be presented later, we actually use the related
quantities $Y^{\eta}_{e,\mu}$, where
\begin{equation}
Y^{\eta}_{e,\mu} \equiv {y^{\eta}_{e,\mu}|_{osc} \over
y^{\eta}_{e,\mu}|_{no-osc}}.
\end{equation}
The numerators are the predictions for $y^{\eta}_{e,\mu}$ in the models,
while the denominators are the same quantities in the absence of
oscillations. We will focus on the case
\begin{equation}
\eta = 0.2
\end{equation}
because our experience with various values for $\eta$ in Ref.\cite{fvy} 
suggests that this choice leads to the largest effects without compromising
too much in regard to statistics.
Using $\eta = 0.2$ also allows comparison of our results with the
preliminary data from SuperKamiokande.

The cases we consider are

\vspace{1mm}

\noindent
(A) Maximal $\nu_{\mu} - \nu_{\tau}$ \cite{barpak} or $\nu_{\mu} - \nu_s$
mixing \cite{epm}.
These two scenarios are indistinguishable as far as up-down
asymmetries are concerned provided that matter effects can
be neglected (which is approximately the case for the
region of $\delta m^2$ considered in this paper). 
The $\nu_{\mu}-\nu_s$ scenario is, in particular,
motivated by the Exact Parity Model in the region of parameter space where
$\nu_e - \nu'_s$ oscillations can be neglected 
for atmospheric neutrinos (Ref.\cite{fvy} focussed on the 
parameter space region where
$\nu_e$ oscillations {\it are} also important). 

\vspace{1mm}

\noindent
(B) Maximal $\nu_{\mu} - \nu_e$ mixing. The Acker-Pakvasa three-flavour model
\cite{ap} is essentially indistinguishable from this scenario for atmospheric
neutrinos. 

\vspace{1mm}

\noindent 
(C) Threefold maximal mixing \cite{giunti}\cite{hps} amongst $\nu_e$,
$\nu_{\mu}$ and $\nu_{\tau}$ in the region of parameter space considered in
Ref.\cite{hps}. 

\vspace{1mm}

\noindent
(D) Massless neutrinos with violation of the Equivalence Principle or
breakdown of Lorentz invariance \cite{gasp}. The special case of exactly
maximal $\nu_e - \nu_{\mu}$ oscillations\cite{fnn}
is considered for definiteness ($\sin^2 2\theta$ can actually be as low as
$0.8$ phenomenologically).

\vspace{1mm}

The Cardall-Fuller three-flavour scheme \cite{cardall} is 
numerically similar to case A with the parameter
choice $\delta m^2 \simeq 0.3$ eV$^2$. For such a large
$\delta m^2$ there are no expected up-down asymmetries  for
the sub-GeV or multi-GeV SuperKamiokande data. 

Consider sub-GeV neutrinos first.  In the water-Cerenkov Kamiokande and
SuperKamiokande experiments, the sub-GeV neutrinos are detected via the
charged leptons $\ell_\alpha$ ($\ell_\alpha$ = $e$ or $\mu$) produced 
primarily from
the quasi-elastic neutrino scattering off nucleons in the water molecules: 
$\nu_\alpha N \rightarrow \ell_\alpha N'$ $(\alpha=e,\mu)$. The event rate is
calculated by integrating the product of the 
differential neutrino flux, the
scattering differential cross-section, the energy 
efficiency function for the detector and
the relevant neutrino oscillation probability with respect to energy and
angular variables (see, for example, Ref.\cite{fvy} for more details). In our
numerical work we have used the differential cross section in Ref.\cite{og}. 
The differential flux of atmospheric neutrinos without geomagnetic effects is
given in \cite{hkkm}, but we have used the differential flux which includes
geomagnetic effects \cite{hkm}. (For other atmospheric neutrino flux
calculations, see Ref.\cite{flux}.) The detector energy efficiency function
can be found in Ref.\cite{kajita1}.

In principle, the charged lepton event rates for multi-GeV neutrinos require
a more involved computation, because several scattering channels must be
considered. Fortunately, as noted in Ref.\cite{lfp}, the up-down asymmetries
are generally insensitive to these details. In practice, all one really needs
to calculate up-down asymmetries reliably is a reasonable estimate for a
function which is effectively a product of the cross-section and the energy
efficiency function. We have estimated this function from the 
SuperKamiokande Monte-Carlo calculations of event rates in the absence of
oscillations. We numerically checked the dependence of the asymmetries on
variations in this ``effective cross-section''. Cases A, B and C vary by only
a few percent, while case D turns out to be uncertain to about the $10\%$
level.

In Figures 1-6 we plot $Y^{0.2}_{e, \mu}$ for each of the cases A, B, C and
D, for three different momentum cuts, as a function of the 
relevant parameter \cite{fn2}. For case A the parameter is 
the difference of squared masses
between $\nu_{\mu}$ and either $\nu_{\tau}$ or $\nu_s$. For case B it is the
$\delta m^2$ between $\nu_{\mu}$ and $\nu_e$. In case C it is the larger of
the two independent $\delta m^2$'s, while for case D it is the usual measure
of either Equivalence Principle or Lorentz invariance violation.  In Figs.\ 1
and 2 we have utilised the standard Kamiokande sub-GeV momentum cuts, $0.2 <
p_{\mu}/GeV < 1.5$ and $0.1 < p_e/GeV < 1.33$, for muons and electrons
respectively. Figures 3 and 4 show the corresponding plots when the
alternative cuts, $0.5 < p_{\mu}/GeV < 1.5$ and $0.5 < p_{e}/GeV < 1.33$, are
employed. Figures 5 and 6 pertain to multi-GeV events. We now discuss some of
the important qualitative features of the graphs: 

Most of these curves show that the asymmetry depends weakly on the neutrino
oscillation parameter ($\delta m^2$ or $\delta v/2$) over a significant range
(almost an order of magnitude). This can be understood as follows:  Neutrinos
with $\cos\theta > 0.2$ travel distances $15 \stackrel{<}{\sim} L/km
\stackrel{<}{\sim} 75$, whereas neutrinos with $\cos\theta < -0.2$ travel
distances $2600 \stackrel{<}{\sim} L/km \stackrel{<}{\sim} 12,700$.  Thus
there will be a large range of $\delta m^2$ or $\delta v/2$ where the
oscillation length is such that neutrinos with $\cos\theta > 0.2$ do not have
time to oscillate, whereas neutrinos with $\cos\theta < -0.2$ oscillate and
are averaged. This weak dependence is interesting because it effectively
provides a prediction within each model for the asymmetries which is
reasonably insensitive to parameter choice and hence avoids the potential
problem of fine-tuning. (Note that the plateau feature in the asymmetries is
also there when $\eta = 0$ is used instead of $\eta = 0.2$. The point is
that, due to the geometry of the situation, the neutrino flux from the
intermediate regime $-0.2 < \cos \Theta < 0.2$ forms a small enough
fraction of the total flux that the qualitative plateau phenomenon persists.)

A comparison of Figs.\ 1 and 3 shows that case A can be more clearly
distinguished from the other cases by adopting the alternative sub-GeV
momentum cut. Case A is in turn clearly different from the no-oscillation $Y
= 1$ case.  Cases B and D can be distinguished from case C through
electron asymmetries, particularly using the alternative sub-GeV cut,
according to Figs.\ 2 and 4. Although case C is not clearly different from
the no-oscillation situation for electron asymmetries, one can return to the
muon asymmetries of Figs.\ 1 and 2 to obtain a clear differentiation. The
most problematic differentiation for the sub-GeV sample is evidently that
between cases B and D. If $\delta m^2$ or $\delta v/2$ is sufficently large
[$> \sim 10^{-2}$ eV$^2$ or (km GeV)$^{-1}$], then this discrimination can be
more easily made.  Fortunately, the multi-GeV cases displayed in Figs. 5 and
6 show a clear difference between cases B and D unless $\delta m^2$ or
$\delta v/2$ is quite small. 

Preliminary atmospheric neutrino results from 
SuperKamiokande can be found in
Ref.\cite{SKprelim}. The asymmetries $Y^{0.2}_{e,\mu}$ can 
be estimated from
the graphs presented for the sub-GeV case with the 
usual momentum cut and for
the multi-GeV case. The results are,
\begin{equation}
Y^{0.2}_e(sub-GeV) \simeq 1.19 \pm 0.10\ (stat.)\qquad
Y^{0.2}_{\mu}(sub-GeV) \simeq 0.80 \pm 0.07\ (stat.)
\label{xx}
\end{equation}
and
\begin{equation}
Y^{0.2}_e(multi-GeV) \simeq 0.80 \pm 0.15\ (stat.)\qquad
Y^{0.2}_{\mu}(multi-GeV) \simeq 0.46 \pm 0.07\ (stat.).
\label{yy}
\end{equation}
These numbers can be compared to Figures 1, 2, 5 and 6. These 
figures show that while case A fits the preliminary data best, the
still significant statistical errors preclude
definitive conclusions. The Cardall-Fuller model fits the data least well.

In summary, up-down asymmetries in the charged leptons induced by atmospheric
neutrinos in the SuperKamiokande detector are important quantities that, with
improved statistics, will be capable of clearly distinguishing 
the various proposed solutions to the atmospheric neutrino anomaly. 

\vspace{1cm}

\centerline{\bf \large Acknowledgements}

This work was supported in part by the Australian Research Council. 
R.F. is an Australian Research Fellow.

\newpage

\centerline{\large \bf Figure Captions}

\noindent
Figure 1.\ \ $Y^{0.2}_{\mu}$ with momentum cut $0.2 < p_{\mu}/GeV < 1.5$
for case A (solid line), B (dashed-dotted line), C (dashed line)
and D (dotted line). 
Note $x = \delta m^2 \ (eV^2)$ for cases A, B and C while
$x = \delta v/2 \ (km^{-1} GeV^{-1})$ for case D.
Also shown (straight dotted lines) is the preliminary
superKamiokande result (Eq.(\ref{xx}) together
with 1 sigma statistical errors.

\vspace{5mm}

\noindent
Figure 2.\ \ 
$Y^{0.2}_{e}$ with momentum cut $0.1 < p_{e}/GeV < 1.33$
for the cases A (solid line), B (dashed-dotted line), C (dashed line) 
and D (dotted line).
Also shown (straight dotted lines) is the preliminary
superKamiokande result (Eq.(\ref{xx}) together
with 1 sigma statistical errors.

\vspace{5mm}

\noindent 
Figure 3.\ \
Same as Figure 1, except that the momentum cut $0.5 < p_{\mu}/GeV < 1.5$
has been taken. 

\vspace{5mm}

\noindent
Figure 4.\ \ 
Same as Figure 2, except that the momentum cut $0.5 < p_{e}/GeV < 1.33$
has been taken. 

\vspace{5mm}

\noindent
Figure 5.\ \ 
$Y_{\mu}^{0.2}$ for the SuperKamiokande multi-GeV sample.
Also shown (straight dotted lines) is the preliminary
superKamiokande result (Eq.(\ref{yy}) together
with 1 sigma statistical errors.

\vspace{5mm}

\noindent 
Figure 6.\ \ 
$Y_{e}^{0.2}$ for the SuperKamiokande multi-GeV sample.
Also shown (straight dotted lines) is the preliminary
superKamiokande result (Eq.(\ref{yy}) together
with 1 sigma statistical errors.

\pagestyle{empty}
\newpage
\epsfig{file=pp1.eps,width=15cm}
\newpage
\epsfig{file=pp2.eps,width=15cm}
\newpage
\epsfig{file=pp3.eps,width=15cm}
\newpage
\epsfig{file=pp4.eps,width=15cm}
\newpage
\epsfig{file=pp5.eps,width=15cm}
\newpage
\epsfig{file=pp6.eps,width=15cm}

\begin{thebibliography}{99}
\bibitem{fn} Note that energy and momentum are almost interchangeable
quantities except for muon events in the 100's of MeV range.
\bibitem{fnn}
In this case the survival probability is given by
$P_{ee} = P_{\mu \mu} = 1 - \sin^2 {\delta v \over 2} E L$
and the transition probability is $P_{e\mu} = P_{\mu e} 
= \sin^2 {\delta v \over 2} E L$ (where $E$ and $L$ are the
energy of the neutrino and the distance travelled respectively).
\bibitem{fn2}
In our numerical work we found some small oscillations of
the $Y$ parameter which we have smoothed out in the Figures
(the oscillations are typically less than 5 percent).
\end{thebibliography}
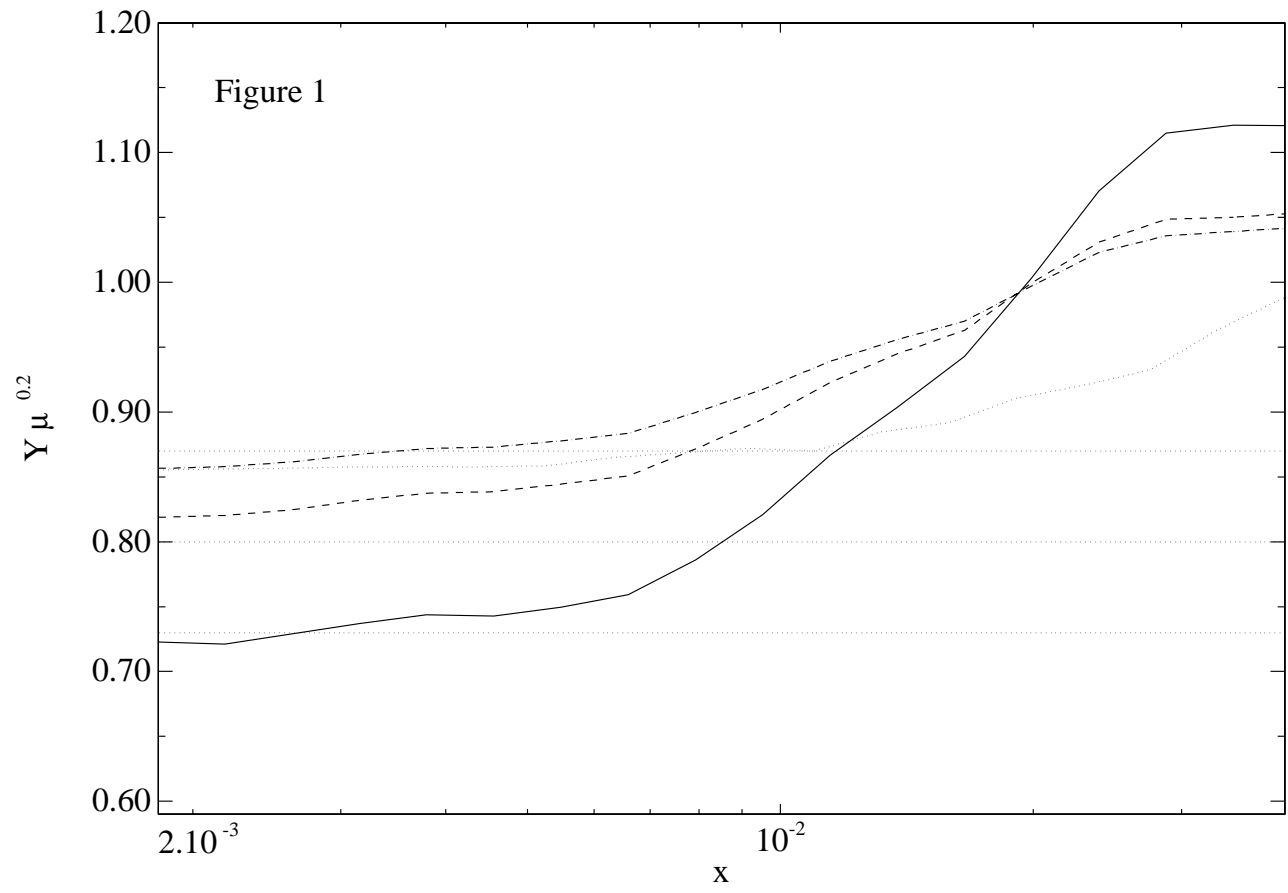
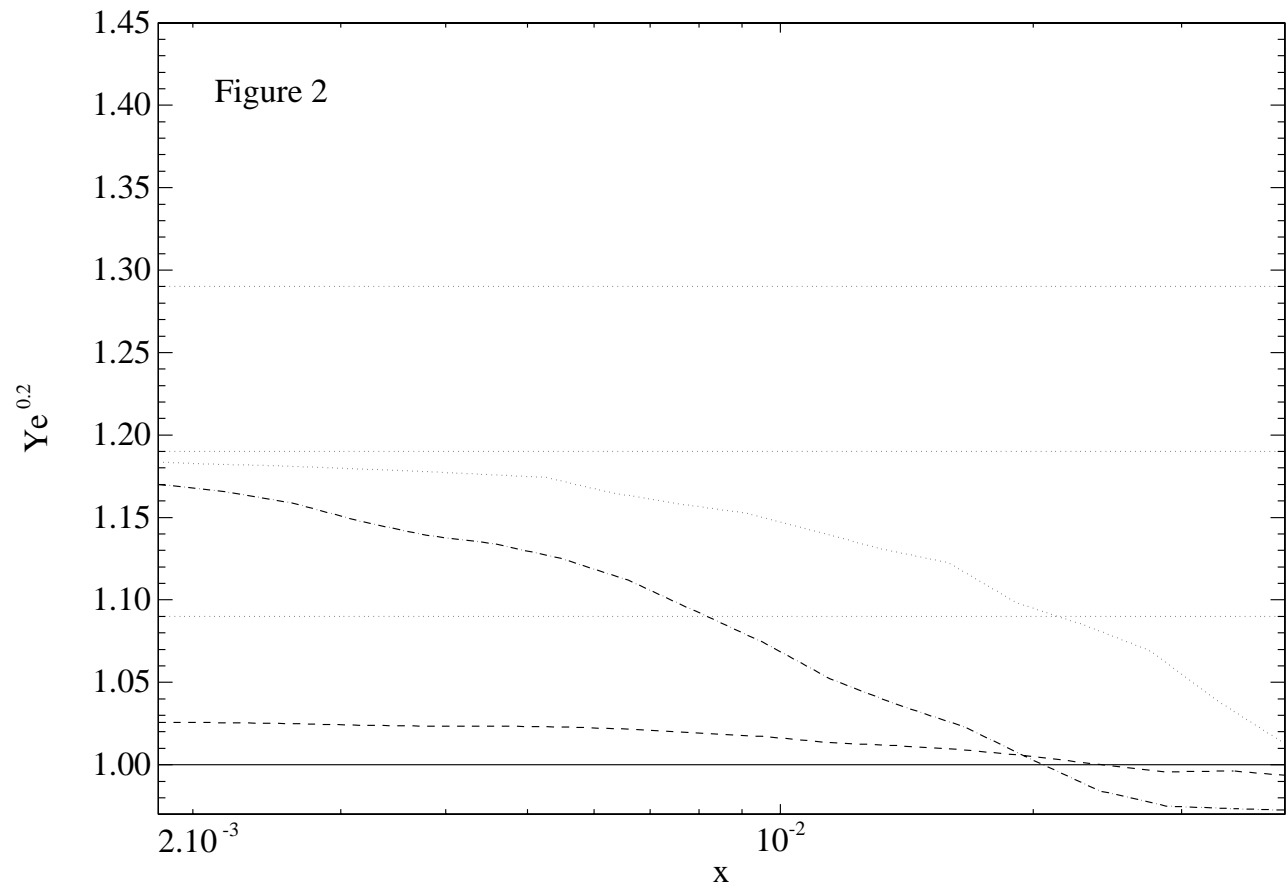
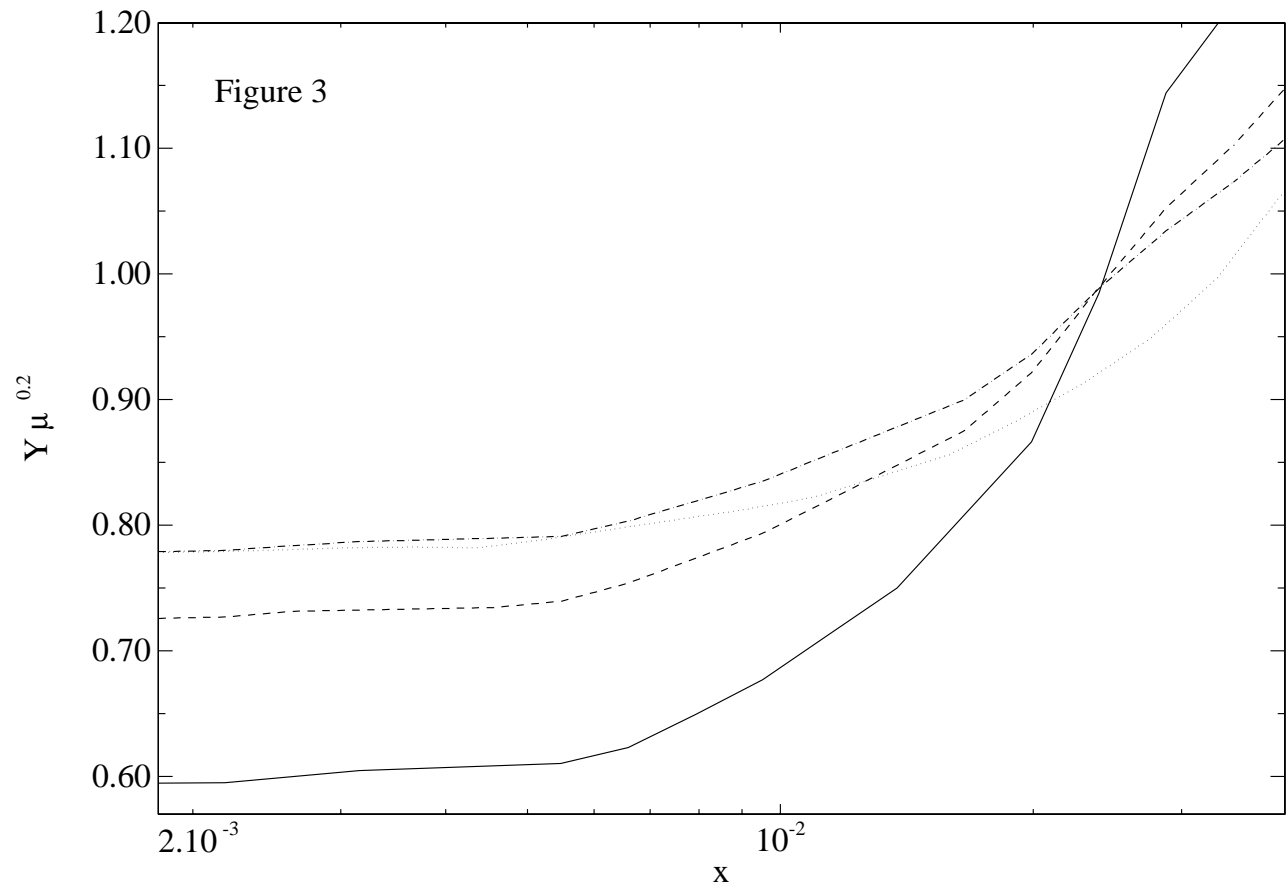
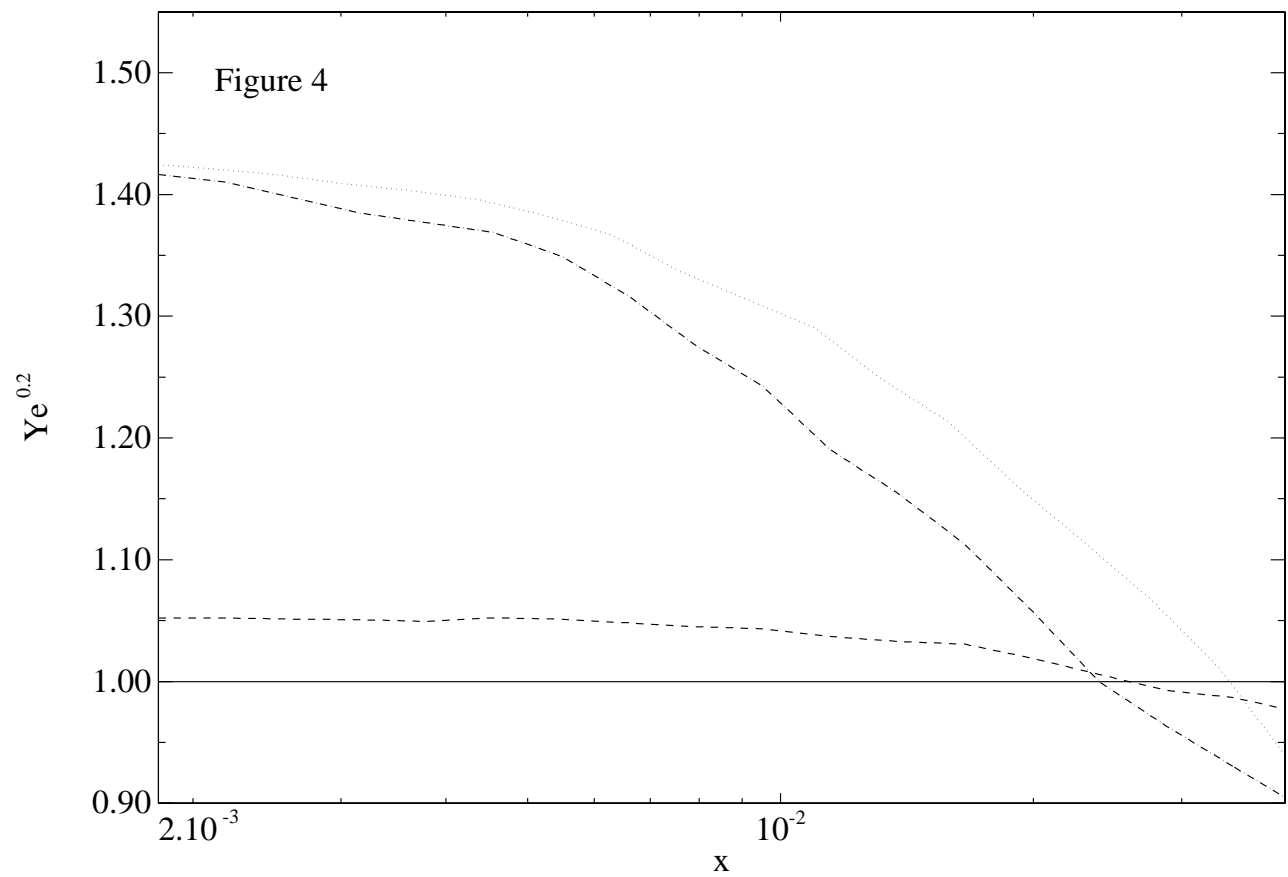
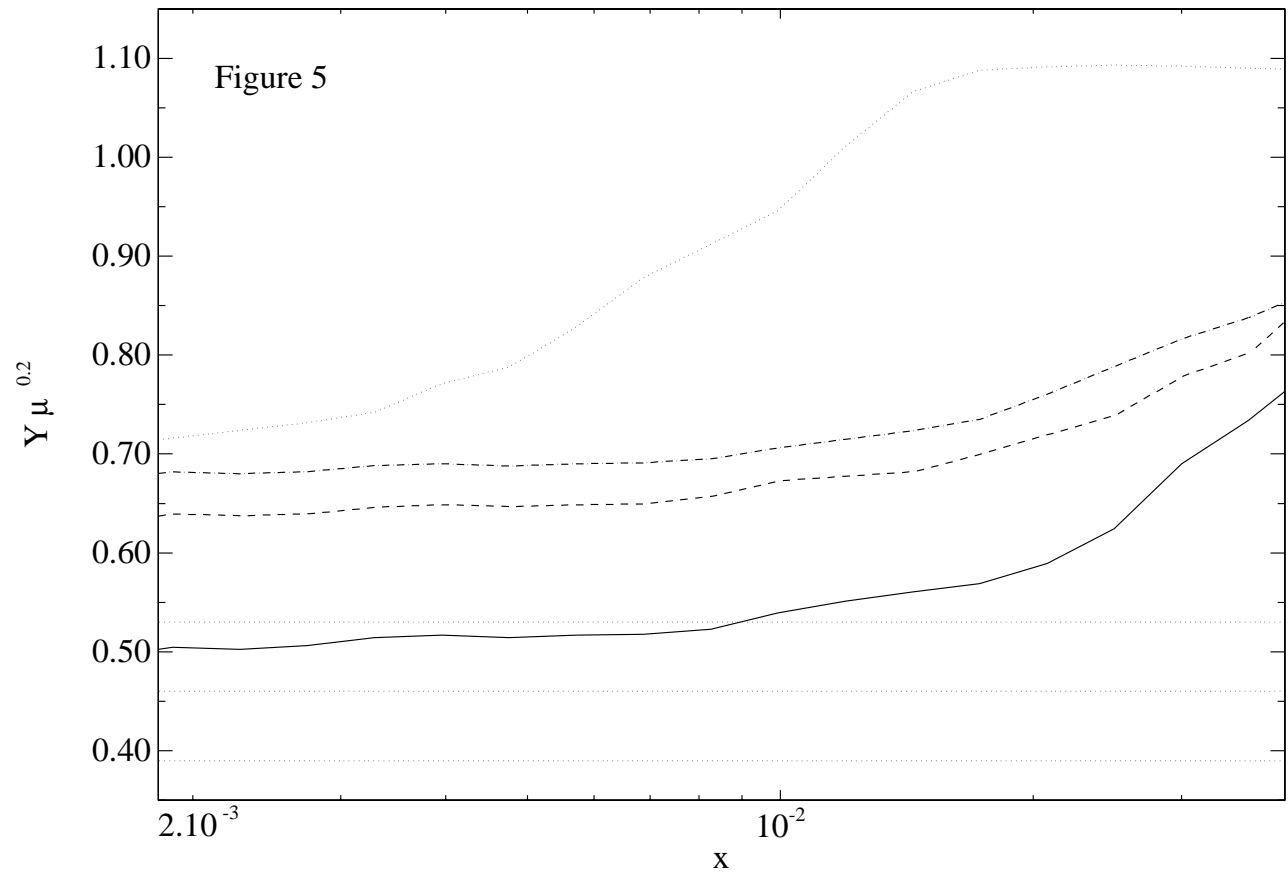
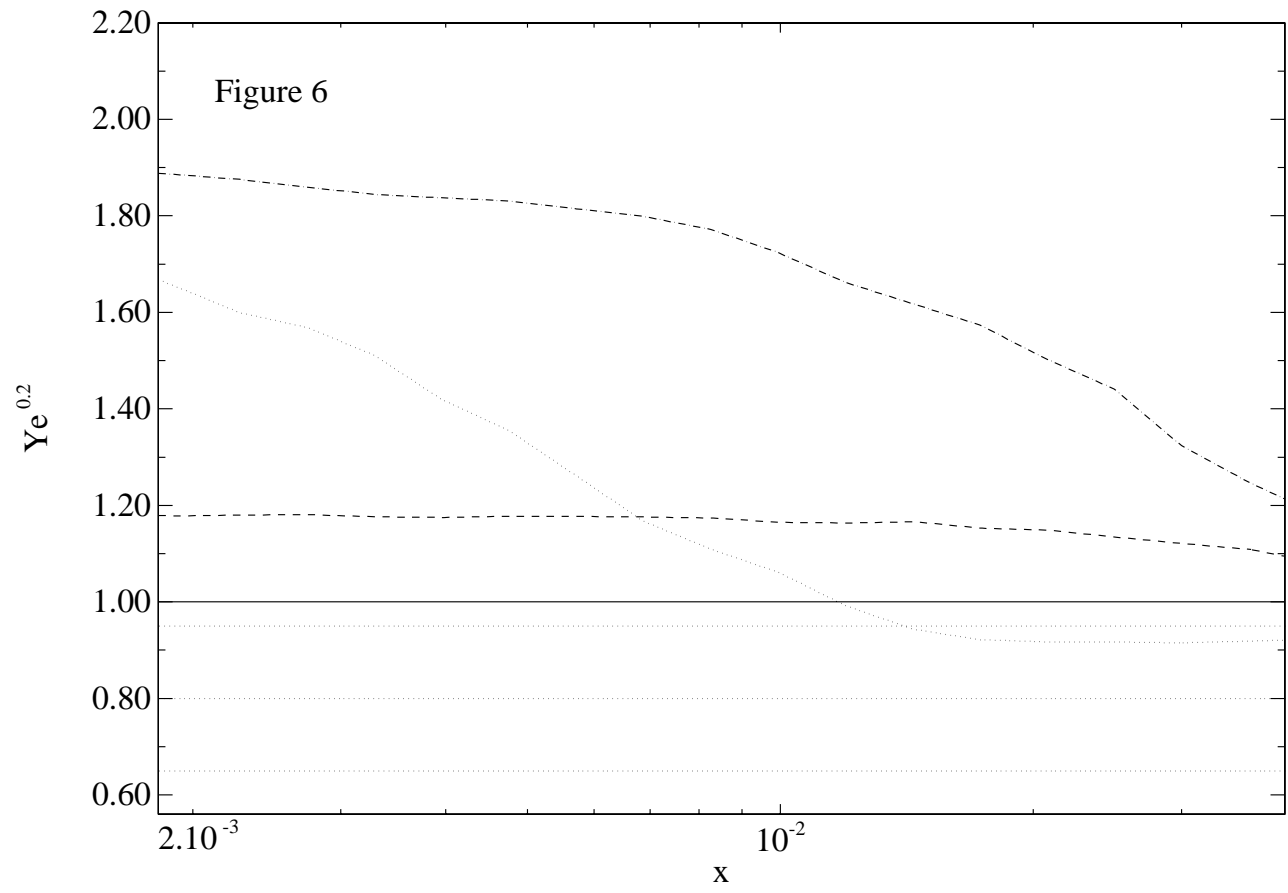
\end{document}